\documentstyle[10pt,aaspptwo]{article}
\def\refpar{ \parindent=0pt\parskip=0.cm\hangindent=1.2em\hangafter=1}
\def\reference{\relax\refpar}

\def \deg {$^{\circ} $}
 \newcount\UGscale   \newdimen\UGdscale
 \newcount\UGleft    \newdimen\UGdleft
 \newcount\UGbot     \newdimen\UGdbot
 \newdimen\UGabs     \UGabs=.01in
\def\UGinsert#1#2#3#4#5#6{
 \dimen0=#3in%
 \vbox to #6\dimen0{
 \vss
   \hbox to \hsize{%
     \hss
     \dimen0=#2in%
     \hbox to #6\dimen0{%
     \UGbody{#1}{#4}{#5}{#6}
     \hss
    }%
    \hss
   }
   \hss
  }
}
 \def\UGbody#1#2#3#4{
    \UGdscale=#4\UGabs \multiply\UGdscale by 100 
    \UGscale=\number\UGdscale \divide\UGscale by \number\UGabs
    \UGdleft=#2\UGabs \multiply\UGdleft by 72
    \UGleft=\number\UGdleft \divide\UGleft by \number\UGabs 
            \advance\UGleft by 5            
    \UGdbot=#3\UGabs \multiply\UGdbot by 72 
    \UGbot=\number\UGdbot \divide\UGbot by \number\UGabs
    \multiply\UGbot by \number\UGscale  \divide\UGbot by 100
    \multiply\UGleft by \number\UGscale \divide\UGleft by 100
    \includegraphics{#1}
    \special {ps::[end] restore }
 }

\begin{document}

\title{VERY HIGH ENERGY GAMMA-RAY EMISSION
FROM THE BLAZAR MARKARIAN 421}

\author{M.~S.~Schubnell\altaffilmark{1},
C.~W.~Akerlof\altaffilmark{1},
S.~Biller\altaffilmark{2},
J.~Buckley\altaffilmark{3},
D.~A.~Carter-Lewis\altaffilmark{4},
M.~F.~Cawley\altaffilmark{5},
M.~Chantell\altaffilmark{3},
V.~Connaughton\altaffilmark{3,6},
D.~J.~Fegan\altaffilmark{6},
S.~Fennell\altaffilmark{6},
J.~A.~Gaidos\altaffilmark{7},
A.~M.~Hillas\altaffilmark{2},
A.~D.~Kerrick\altaffilmark{4,8},
R.~C.~Lamb\altaffilmark{4},
D.~I.~Meyer\altaffilmark{1},
G.~Mohanty\altaffilmark{4},
J.~Rose\altaffilmark{2},
A.~C.~Rovero\altaffilmark{3,9},
G.~Sembroski\altaffilmark{7},
T.~C.~Weekes\altaffilmark{3},
C.~Wilson\altaffilmark{7},
and J. Zweerink\altaffilmark{4}}

\altaffiltext{1}{University of Michigan, Ann Arbor, MI 48109}
\altaffiltext{2}{University of Leeds, Leeds LS2 9JT, UK}
\altaffiltext{3}{Whipple Observatory, Harvard-Smithsonian CfA,
Amado, AZ 85645}
\altaffiltext{4}{Iowa State University, Ames, IA 50011}
\altaffiltext{5}{St.~Patrick's College, Maynooth, Co. Kildare, Ireland}
\altaffiltext{6}{University College Dublin, Belfield, Dublin 4, Ireland}
\altaffiltext{7}{Purdue University, West Lafayette, IN  47907}
\altaffiltext{8}{Postal Address: Department of Physics and Geology,
Northern Kentucky University, Highland Heights, KY 41099}
\altaffiltext{9}{Postal Address: Instituto de Astronom\'\i a  y  F\'\i sica
del Espacio, C.C. 67, Suc. 28, 1428 Buenos Aires, Argentina}

\begin{abstract}
Very high energy gamma-ray emission from the BL Lac object Markarian
421 has been detected over three observing seasons on 59 nights
between April 1992 and June 1994 with the Whipple 10-meter imaging
Cherenkov telescope.  During its initial detection in 1992, its flux
above 500 GeV was 1.6$\times$10$^{-11}$photons cm$^{-2}$ s$^{-1}$.
Observations in 1993 confirmed this level of emission.  For observations
made between December 1993 and April 1994, its intensity was a factor of
2.2$\pm$0.5 lower.  Observations on 14 and 15 May, 1994 showed an
increase over this quiescent level by a factor of $\sim$10 (Kerrick et
al. 1995).  This strong outburst suggests that 4 episodes of increased
flux measurements on similar time scales in 1992 and 1994 may be
attributed to somewhat weaker outbursts.  The variability of the TeV
gamma-ray emission from Markarian 421 stands in contrast to EGRET
observations (Lin et al. 1994) which show no evidence for
variability.
\end{abstract}

\keywords{gamma rays: observations; blazars: individual: Markarian 421.}

\twocolumn

\section{Introduction}

The central engines of active galactic nuclei (AGN) are presumed to be
massive black holes whose accretion powers radiation over a wide range
of the electromagnetic spectrum. More than a score of these have now
been reported as sources of multi-hundred MeV/GeV photons (Fichtel et
al. 1994), all of which are identified with the sub-class of AGN which
are radio-loud and core dominated, and most of which are blazars. (See
Dermer \& Schlickeiser 1992, for a glossary of terms relevant to
AGN).

The AGN reported by EGRET vary in brightness by a decade and range in
redshift from 0.031 to above 2. The nearest such object is Markarian
421, a BL Lac object extensively observed at radio (Owen et al. 1978;
Zhang \& Baath 1990), UV/optical (Maza, Martin, \& Angel 1978;
Mufson et al. 1990), and X-ray frequencies (Mufson et al. 1990;
Mushotzky et al. 1979; George, Warwick, \& Bromage 1988).  Its
parent galaxy has been identified as a giant elliptical (Ulrich et
al. 1975; Mufson, Hutter, \& Kondo 1989) and it was the first BL
Lac established as an X-ray source (Ricketts, Cooke, \& Pounds 1979).
Using VLBI techniques, the radio fine structure of the source
has been mapped with a resolution of 1 mas to show a core-jet-like
structure which exhibits possible superluminal motion (Zhang \&
Bhaath 1990). Variability with a time scale of hours has been
observed in X-rays (Giommi et al. 1990) and at
lower energies (Xie et al. 1988).
Historically the source has exhibited low amplitude
variability on the scale of months to years in the
radio emission (Aller \& Aller 1995) but so far there is no evidence for
variability in the gamma-ray emission in the high
energy regime (100 MeV - 10 GeV, Lin et al. 1994).
   
A comprehensive summary of recent theoretical work relevant to the
blazars is given by von Montigny et al. (1995).  Although the ultimate
origin of the blazar power is the central engine, high energy gamma-ray
emission is likely to be beamed from a jet of highly
relativistic particles.  (If one assumes isotropy for the emission,
enormous luminosities, in some cases more than 10$^{49}$ erg/s,
would be required.)  Furthermore any gamma rays at GeV-TeV energies
that may be created in the inner regions of the source would be
readily absorbed via the pair production process.  Because of the
likely jet geometry associated with the high energy emission process,
time variations observed at a single energy, if present, do not place
strong constraints on the region of emission since there are many ways
in which that variability may arise.  However, correlated time scale
variations at different energies (e.g. hard synchrotron and TeV) are
generally useful and are the predicted consequence of models in which
the gamma rays are produced by inverse Compton scattering.
      
It has been pointed out by Gould and Schr\'eder (1967) and recently
reformulated by several authors
(Steck\-er, De Jager, \& Salamon 1992; Salamon, Steck\-er
\& De Jager 1994; Dwek \& Slavin 1994; MacMinn \& Primack 1995)
that the interaction of gamma rays with intergalactic infrared and
optical photons produced by galaxy formation may severely attenuate
the TeV photon emission from cosmologically distant objects and that
evidence for such absorption (or lack thereof) may be used to
constrain both the energy density of starlight and the distance of the
sources. The dynamic energy range covered by the Whipple Observatory
instrument is particularly sensitive to such effects for the assumed
distance to Markarian 421 (124 Mpc, H$_0$=75 km s$^{-1}$ Mpc$^{-1}$)
and an upper limit on the IR density field from our observation is
derived in a separate paper (Biller et al. 1995).

\section{Instrument and Observations}

The observations reported in this paper were carried out with the very
high energy gamma-ray telescope (Cawley et al. 1990) at the
Whipple Observatory on Mt. Hopkins in southern Arizona (31$^{\circ}$
41' north, 110$^{\circ}$ 53' west, alt = 2300 m).  This 10 meter
diameter telescope images Cheren\-kov light from air showers on a
hexagonal array of 109 fast photomultiplier tubes (PMT's) in the focal
plane.  The inner 91 2.86 cm diameter tubes of this high resolution
camera have a 0.25$^{\circ}$ spacing. A surrounding outer ring of 18
5.08 cm diameter PMT's was replaced in August 1993 with a partial ring
of 18 2.86 cm tubes.
 
The effective energy threshold $E_{\it th}$, above which there is
100\% detection efficiency over the effective collection area, was
determined to be 0.4 TeV by simulations for the 1988/89 observing
season.  For later periods, an adjustment due to changes in mirror
reflectivity and camera configuration was made by calibrating with the
measured cosmic ray rate {\it R} assuming a scaling behavior
as {\it R} $\propto E_{th}^{-1.7}$ for the energy threshold $E_{th}$. 
An energy threshold of 400 GeV corresponds to an average count rate of
270 min$^{-1}$.  This scaling method can also be used to estimate
gamma-ray event rates at different energy thresholds for a source with
a power-law spectrum $E^{-\alpha}$ and known spectral index $\alpha$.
In the context of this paper, we assume for the photon emission from
the Crab Nebula an integral energy spectrum of the form $E^{-1.4}$ as
derived by Vacanti et al. (1992)

A shower image is recorded when at least $m$ out of the inner 91 tubes
exceed the hardware threshold of $n$ photoelectrons. With the
standard trigger condition ($m$=2, $n$=50) a typical trigger
rate of 4-5 Hz with $E_{th}$=400 GeV is obtained.  All the observations were
restricted to zenith angles less than {40\deg} where, on the basis of
our Crab Nebula observations, the rate variation with zenith angle is
smaller than 10\%.  More detailed descriptions of the instrument and
the Cherenkov light imaging technique can be found elsewhere (Cawley
et al., 1990, Lewis 1990, Reynolds et al., 1993).  The majority of the
data were
taken in a standard on/off tracking mode in which the source region
and a background region are observed for equal times (usually 28
minutes).  The location of the background region differs from that of
the source region only in right ascension; thus both regions have the
same zenith angle coverage. This procedure eliminates systematic effects
introduced by the small variation in counting rate due to zenith angle changes.
  
For a well established source such as Markarian 421, the effective
observing time can be doubled by observing in an alternative mode in
which the telescope always follows the source (\lq tracking mode').
Under these conditions, shower selection procedures, described below,
are used to establish an appropriate value for the rate of background
showers.  Table 1 summarizes the varying detector parameters for the
three observing periods between 1992 and 1994.

Data were processed following the standard Whipple analysis procedures
(Reynolds et al. 1993) in which the Cherenkov light images are
flatfielded, cleaned and characterized by simple image moment
parameters (Hillas 1985).  The shape of the approximately elliptical
shower image and its location relative to the assumed source location
in the image plane are specified by the parameters {\it width, length}
and {\it distance}. The fourth parameter, {\it alpha}, is defined to
be the angle between the major axis of the shower image and a line
from its centroid to the assumed source location in the image plane.
For gamma-ray showers from a point-like source, {\it alpha} should
be near 0$^{\circ}$ since the elliptical images point to the location
of the source in the image plane. 

Information on the shape and the orientation of the images is used to
remove more than 99\% of background cosmic-ray showers, while keeping
a substantial fraction $\ge 0.5$) of possible gamma-ray showers.  For
this analysis we applied the same discrimination values as we have for
the selection of gamma-ray events from the Crab Nebula (0.073\deg
$ < ${\it width}$ < $ 0.150\deg, 0.16\deg $ < ${\it length}$ < $ 0.3\deg,
0.51\deg $ < ${\it distance}$ < $ 1.1\deg, {\it alpha}$ < $ 15\deg).  This
image selection procedure (``supercuts'') was successfully applied to
repeated observations of the Crab Nebula and has proved to be the most
sensitive and reliable technique so far developed to discriminate
gamma-ray events from the dominant hadronic background (Fegan et
al. 1994).

Motivated by the detection of Markarian 421 by the EGRET instrument at
GeV energies (Lin et al. 1992), observations with the 10-meter
reflector were made initially between March 24 and June 2, 1992 for a
total of 445 minutes on-source and an equal amount of time off-source
during 8 clear nights.  With the instrument operating at an effective
energy threshold of 500 GeV, we obtained strong evidence for gamma-ray
emission from the direction of the source (Punch et al. 1992). 

In the fall of 1992, several instrumental improvements were made
including the recoating of most of the mirror facets of the 10-meter
reflector.  Preparations for operation in coincidence with the nearby
new 11-meter telescope (Schubnell et al. 1992) resulted in changes in
the trigger conditions as well. These modifications had the effect of
changing the optimum selection criteria used for selecting gamma-ray
images and made it necessary to introduce an image size selection in
software to adjust the energy thresholds similar to that used
previously. This selection, which excludes images with a
total number of less than 350 photo electrons, was chosen to reproduce
the expected Crab gamma-ray rate at 500 GeV and not to optimize the
signal.
During the following winter 1992/1993, the AGN observation program was
continued and from December 23, 1992 to February 26, 1993, Markarian
421 was again observed on 10 clear nights for a total of 750 minutes.

For the most recent observations between December 16, 1993 and June
12, 1994, the mirror re-coating was completed and the trigger condition
reverted to that of previous years.  The instrument was operating at
an energy threshold of 350 GeV, somewhat lower than in 1988/1989 due
to the higher mirror reflectivity.  Prior to the observations in May
1994, we installed light focussing cones in front of the PMT plane
(Kerrick et al. 1995).
This increased the number of Cherenkov photons collected by the
photomultipliers and decreased the trigger energy threshold.
Observations with the light cones in place produced stable trigger
rates and first measurements on the Crab Nebula showed a significant
increase in the observed gamma-ray rate. Encouraged by these results, we
have continued to use the light cones throughout the entire observing
season as well as in 1994/1995 with reproducible results on the Crab
Nebula, consistent with a 250 GeV energy threshold.

For the purpose of calculating the photon flux values and to
investigate possible time variations, we reference all fluxes to the
observed intensity of the Crab Nebula.  This procedure corrects for
possible systematic variations which are due to the different trigger
conditions over the entire observing period, assuming the uniformity
of the Crab emission.  An analysis of the 4-year database of Crab Nebula
observations shows that the flux is constant to better than 10\%
(Weekes et al. 1993). This limit of variability is included as a
systematic error of 10\% to the calculated Markarian 421 fluxes.

\section{Results}

Because of the changes in the instrument as discussed in the previous
section we will treat each of the observing seasons separately.  The
sensitivity of the Whipple telescope is such that a significant flux
of 0.3 of the Crab flux can be observed from a single night's
observations. Therefore the dataset has been divided into daily
intervals, with each interval spanning a minimum of 1/2 hour
observation. A summary of the observations for the three seasons
reported here can be found in table 2.

\subsection{The 1992 Season}

The first observation of very high energy gamma-ray emission from
Markarian 421 was based on data taken during the 1992 March-June
period (Punch et al. 1992). We re-analyzed this dataset using the
standard image analysis procedure but slightly modified the algorithm
to calculate the relative photomultiplier gains and excluded phototubes
for which the pedestal distributions were outside the
statistical expectation.  With the definition of the gamma-ray signal
as $(N_{on}-N_{off})$, where $N_{on}$ represents the number of
selected images according to the discrimination procedure discussed in
the previous section in the on-source scan and $N_{off}$ represents
the number of selected images in the off-source observation, we obtain
an average gamma-ray rate of 0.34$\pm$0.05 min$^{-1}$ over the total
445 minutes of observation. With the statistical significance $\sigma$
defined as $\sqrt{N_{on}+N_{off}}$, we find that the observed excess
has a significance of $6.7 \sigma$.  

Figure 1 a) shows how the difference $N_{on}-N_{off}$ is distributed
as a function of the orientation parameter $alpha$ after applying all
other selection criteria. The excess at small $alpha$ values is expected
for gamma-ray events from the source direction, while hadron-initiated
Cherenkov events are random in orientation and contribute an
isotropic background.  Assuming an effective area of $3.5 \times
10^{8}$ cm$^{2}$, the measured rate is equivalent to an integral
gamma-ray
flux of 1.59 $\pm$ 0.20 $\times$ 10$^{-11}$ cm$^{-2}$ sec$^{-1}$
above 500 GeV (see table 3).

Two stars (SAO 62387 and SAO 62393) with a combined brightness
equivalent to magnitude 5 are located near the Markarian 421 position.
This led us to investigate the possibility of a false signal due
to the brightness variation in ``on'' and ``off'' fields.
The angular separation from Mrk 421 is on the order of the telescope's
optical resolution and thus both star images are always contained in
the center tube.
The most obvious test is to exclude the photomultiplier tube
containing these stars (the center tube) from the analysis. No significant
difference in the result was found. When we excluded the center tube from the
trigger and the analysis during observations made in the following years,
we also could find no evidence for a systematic effect.
Additional control observations on two stars of similar
magnitude show null results when subjected to the supercuts analysis.

Figure 2 a) shows the emission on a nightly basis for the 1992
observing period.  The errors shown are statistical only and do not
incorporate small systematic effects due to the variation in zenith
angle of the observations.  The maximum deviation (2.9 $\sigma$) from
a steady flux is found on MJD 48764 (1992 May 22 UT). A $\chi^2$ test of
the 1992 data gives a 2\% probability that the observed flux is
consistent with constant emission (see table 4).

If we exclude the observation from MJD 48764 from the sample and repeat
the $\chi^2$ test on the Markarian 421 data, the probability for
constant observed emission then increases from 2\% to 40\%. We
emphasize that the observed increase in photon emission on MJD 48764
has a marginal significance but, in context with the later observed
day-scale flux variations, has to be considered as a likely flare. The
apparent photon flux for this particular 38 minute observation is
5.0$\pm$ 1.2 $\times$ 10$^{-11}$ cm$^{-2}$ sec$^{-1}$ above 500 GeV.

To exclude possible systematic effects in the instrument as the source
for the variability in the emission, we have analyzed contemporaneous
data from the Crab Nebula (Figure 3 a)).  From this we extract
a gamma-ray rate of
0.80 $\pm$ 0.08 min$^{-1}$ corresponding to an integral photon flux of
3.81$\pm$ 0.03 $\times$ 10$^{-11}$ cm$^{-2}$ sec$^{-1}$ above 500 GeV.
A $\chi^2$ test for constant flux from the Crab Nebula gives a 48\%
probability.

\subsection{The 1992/1993 Season}

For the 1992/1993 season, changes in the trigger conditions (Table 1)
resulted in a loss in effective collection area in spite of the fact
that a large fraction of the mirrors had been recoated.  In order to
have a consistent database for Markarian 421 for 1993, we have used
only that data for which contemporaneous Crab observations under
identical instrumental conditions are available as reference.  Two
instrumental effects had to be considered for the data analysis: 1) a
new PMT camera with slightly different tube spacing was installed
and 2) a persistent problem with one tube led to a software filtering
procedure to remove triggers associated with that tube.

Following the above selection, a total of 750 minutes of on-source
observations of Markarian 421, distributed over 10 nights, was
obtained.  Applying the standard analysis, we obtain an
average gamma-ray
rate of 0.29 $\pm$ 0.08 min$^{-1}$ with a significance level of
3.7 $\sigma$ above 400 GeV. 

From the observations of the Crab Nebula, we determined that a large
portion of the trigger-induced background can be suppressed if we restrict
the analysis to images with a total of at least 350 photoelectrons.
This selection increases the effective energy threshold to
500 GeV.  The corresponding on$-$off $alpha$ distribution (with the 350
pe cut applied) is shown in figure 1 b). 

Although we have almost doubled the observing time compared to the
1992 season, the significance in the extracted signal is considerably
lower due to the increased contamination of the selected gamma-ray
sample with hadronic showers (Signal/Background = 0.29 compared to
Signal/Background = 0.85 for the 1992 observations).  The observed
average gamma-ray rate is equivalent to an integral flux of $1.67 \pm
0.4 \times 10^{-11}$ cm$^{-2}$ sec$^{-1}$ above 500 GeV, in good
agreement with the result from the previous year if we scale the 400
GeV 1992 result to 500 GeV, assuming an integral spectrum like
$E^{-1}$.  (The assumption of an integral spectrum $E^{-1}$ for
Markarian 421 in this context is justified by the extrapolation of
the (integral) 1992 flux to the lower energy EGRET data points (Lin et
al. 1994).) 

The nightly flux values were calculated to test for variability of the
emission (Fig 2b).  The maximum deviation ($1.3 \sigma$) from a steady
flux is found on MJD 49009 (1993 Jan 22 UT). A $\chi^2$ test
gives a 84\% probability
that the flux is constant (table 4).

\subsection{The 1994 Season}

In an effort to monitor more closely the photon emission from
Markarian 421, extensive observations were taken during the period
December 15, 1993 - June 12, 1994 (MJD 49336 - MJD 49515).
The instrument operated with two
slightly different energy thresholds, depending on whether or not
light focussing cones were in place.  All the scans prior to May 10th
were taken without the light cones with the exception of observations
on January 17 (MJD 49369), March 5 (MJD 49415) and March 6 (MJD 49416).
For the runs without the light cones,
the energy threshold is already significantly lower ($E_{th}$=350 GeV)
than for the previous year, an improvement due to the higher
reflectivity of the newly coated mirrors.  The energy threshold for
observations with the light cones is $E_{th}$=250 GeV.

During the night of May 15 (MJD 49487) an increase 
by a factor of nine in the Markarian 421 TeV emission was
observed (Kerrick et al. 1995a).  Because
of the occurrence of this outburst we separate the observations into
pre-burst, burst, and post-burst periods.
\medskip

a) pre-burst

For observations in the standard on/off mode between December 1993 and
April 1994 we measured an average gamma-ray rate of 0.31 $\pm$ 0.06
min$^{-1}$ during 1145 minutes, on 16 nights with a significance of
5.5 $\sigma$ (Table 2). The energy threshold during this period is
calculated from contemporaneous Crab observations ($R_{Crab}$ = 1.89
min$^{-1}$) to be 350 GeV. In addition, we obtained observations without
the light cones in tracking mode on 9 nights totaling 939 minutes
resulting in an average gamma-ray rate of 0.17 $\pm$ 0.06.  During 4
nights on which light cones were in place, we collected an almost equal
amount of tracking mode and on/off mode data (Table 2).  The average
observed gamma-ray rate is 1.59 $\pm$ 0.18. From table 2 and figure 2c
it can be seen that this high average rate is due to an increase in
the observed emission on March 5 and March 6. For the night of March 6,
we have observations of the Crab Nebula which we analyzed to exclude
the possibility of systematic effects which could be attributed to
either the newly mounted light cones or the non-standard observing
technique (tracking) for those two nights.  On March 6, 1994, we
observed the Crab Nebula for 29 minutes under identical instrumental
conditions as for the Markarian 421 observations. The detected
gamma-ray rate is 2.9 $\pm$ 1.1 min$^{-1}$, in agreement with the
average of 3.35 $\pm$ 0.39 min$^{-1}$ for all the Crab Nebula scans at
the 250 GeV energy threshold. In the absence of any other systematic
effect (for example unstable weather conditions) the observed excess is
statistically significant. A $\chi^2$ test
applied to all pre-burst data gives a probability of 1.3 $\times$
10$^{-9}$. The largest contribution comes from March 6 (see figure 2c).
The average gamma-ray rate for the pre-burst period (excluding March
6) for the combined data between December 15, 1993 (MJD 49336)
and May 9, 1994 (MJD 49481) is
0.23 $\pm$0.04 min$^{-1}$ above 350 GeV. If we scale from the average
of the 1992 and 1993 observing season (0.32 min$^{-1}$), we expect
a rate of 0.50 min$^{-1}$ above 350 GeV. Even if we allow an
additional systematic error of 10\%, the constant emission from
Markarian 421 between December 1993 and May 1994 is,
by a factor of 2.2$\pm$0.5,
significantly below that of the two previous observing seasons.
\medskip

b) burst

While on May 9th (MJD 49419), Markarian 421
was still observed at the low flux
level of the previous months, a significant increase of the photon
flux was observed on May 15th (MJD 49487, see fig 2c), reaching a maximum of
2.1$\times$10$^{-10}$ cm$^{-2}$ s$^{-1}$ during a 28 minute scan
(Kerrick et al. 1995a).  During this observation we measured a
gamma-ray event rate at the level of 4.5$\pm$0.8 min$^{-1}$, a rate
comparable to the Crab at this energy level.  The appearance of the
Moon during the following nights made observations of Markarian 421
impossible again until May 30 (MJD 49502).  The TeV outburst
of Markarian 421 occurred during a multiwavelength campaign and simultaneous
or contemporaneous observations are available.  The strongest observed
TeV flux precedes the observation of an increase in the hard X-ray
emission by the ASCA satellite by 24 hours (Takahashi et al. 1994).
Simultaneous observations by the EGRET instrument did not indicate an
increase in the GeV emission. A detailed analysis of the
multiwavelength data is discussed in a separate paper (Macomb et
al. 1995).
\medskip

c) post-burst

Following the outburst in May, we monitored Markarian 421 during the
month of June when permitted by weather and moon. All observations were
made in the standard on/off mode with light cones in place. The average
observed gamma-ray rate is 0.43 $\pm$ 0.13 min$^{-1}$ over the course of
10 nights (see figure 2c) with a probability of 11\% for
constant emission (see table 4).
                   
The average integral photon fluxes measured at 250 GeV and at 500 GeV
are consistent with the extrapolation of the spectrum measured between
100 MeV and several GeV (see figure 4). As of the time this paper is
written, no detection of gamma rays from Markarian 421 has been claimed
by experiments sensitive to gamma-ray energies above 10 TeV.  Reported
upper limits on the photon emission from those experiments
(Alexandreas et al. 1993, Amenomori et al. 1994, Catanese et al.
1994, Karle 1994, K\"uhn 1994) have been incorporated in figure 4.

\section{Discussion}

\begin{sloppypar}
The average gamma-ray flux observed during 1992 and 1993 corresponds
to a photon luminosity of $\sim 10^{43}$ erg s$^{-1}$ above 500 GeV
assuming isotropic emission at a distance of 124 Mpc.  However, the
fact that all identified extragalactic sources detected by EGRET can
be associated with the blazar class of AGN suggests that the GeV
emission is highly beam\-ed, originating in the relativistic plasma
outflow which forms the radio jets. Beamed emission of the gamma rays
furthermore overcomes the strong photon-photon absorption implied by the
large gamma-ray luminosities for the case of isotropic emission. If we adopt
the picture of beamed emission for the TeV component, the true 
VHE gamma-ray luminosity decreases by a factor
of $\sim$ 10$^{-3}$ for beam opening angles of the order of a few
degrees.

\end{sloppypar}

Much attention has been devoted lately to the discussion of possible
absorption of the TeV photons in the intergalactic radiation field
(Stecker, De Jager, \& Salamon, 1992, Salamon, Stecker, \& De Jager,
1994) i.e. gamma rays with energies of 0.5 TeV - 5 TeV interacting with
soft 0.05 - 0.5 eV photons to produce e$^+$e$^-$ pairs.  For a
relatively nearby source like Markarian 421 (z=0.031) this is a rather
small effect.  (The possible presence of an absorption effect in our
original 1992 data (Mohanty et al. 1993) has been used by De Jager et
al. (1994) to determine a value for the infrared energy density.
A more conservative approach has been adopted by Biller et al. (1995)
to establish an upper limit to the energy density.)

The question arises whether or not this process could be occurring in
the vicinity of the AGN, that is to say, can gamma rays with TeV
energies escape the photon field close to the central engine of the
AGN?  For some models this requirement is a severe obstacle as
explained for instance by Coppi, Kartje, \& K\"onigl (1993). While
their theory nicely models the non-thermal emission component by
upscattering radiation (ambient or synchrotron-self-Compton produced)
by a relativistic
electron/proton beam, they need to place the TeV emission regime
further from the core than the GeV emission in order to avoid
absorption in the region of high energy density at small distances.
In simultaneous observations during the May 1994 TeV outburst
(Macomb et al. 1995), the EGRET instrument did not observe an increase
in the photon emission from Markarian 421 above the previously detected
level (Lin et al. 1994).

Many of the various models that have been proposed to explain the origin
of the gamma-ray emission favor production of the gamma-rays by the
inverse-Compton process. Beamed relativistic electrons in the jet upscatter
low energy photons from synchotron radiation (Ghisellini \& Maraschi, 1989;
Marscher \& Bloom, 1992) or
other ambient sources (Blandford 1993; Dermer, Schlickeiser,
Mastichiadis, 1992; Sikora, Begelman, Rees, 1994)
to gamma ray energies. These models imply a strong correlation between the
production of radiation across the non-thermal waveband. Any change in the
electron flow or electron distribution translates into a variation in the 
emission at other wavelengths. Differences in
the relation of variability at given frequencies
reflect distinct production regions.
The
inhomogeneous synchrotron-self-Compton
model by Maraschi, Ghisellini, and Celotti (1992) for instance
places the production of gamma-rays at the
inner part of the jet while hard X-rays (as well as radio and IR emission) are
produced in the outer jet part and thus 
fluctuations in the gamma-ray
band are predicted to lead X-ray variability by typically a day.

In a similar model by Sikora, Begelman, and Rees (1994), the entire
gamma-ray spectrum is produced within the same jet region through
Comptonization of ambient radiation (from dust near the jet) by
relativistic electrons. The authors point out that the problem of TeV
opaqueness can be overcome if the seed photons come from thermal IR
emission by dust rather than UV photons, supported by the fact that
while thermal UV excess is common for OVV quasars, it is only observed
in very few BL Lac objects.
This hints that the non-observation of EGRET detected sources other
than Markarian 421 (Kerrick et al. 1995b)
above a few hundred GeV may be a
combination of an intrinsic source feature (BL Lac's vs. OVV's for
example) and the increasing IR absorption towards higher
redshifts. Clearly, this needs more observational evidence, some of
which can be provided by future measurements of AGN emission spectra
in the (10-100) GeV regime where both processes will potentially cut off
the power-law spectrum observed at high energies.
  
In contrast to the leptonic models is the approach by Mannheim
(Mannheim 1993; Mannheim \& Biermann 1992) in which a shock
accelerated ultra-re\-la\-tivis\-tic proton population in the jet
photoproduces $\pi^0$'s and the observed gamma rays are then
produced by synchrotron
cascade reprocessing. In this model, the gamma-ray emission observed at
GeV energies extends naturally into the TeV regime. No predictions
about the correlation of variability in the TeV gamma-ray emission to
the GeV and X-ray emission are made but it was pointed out by von
Montigny et al. (1995) that a disturbance in the proton population can
translate to large gamma-ray flares downstream from the cascade origin.
 
Although the Whipple collaboration has recently
reported a somewhat weaker TeV AGN source, Markarian 501 (Quinn et al. 1995),
Markarian 421, is, so far, the best probe to test models at the highest
accessible energies.
The detection of a TeV component in the Markarian 421
emission as well as the observation of day scale variability already
constrains models of gamma-ray production in
relativistic jets. Additional,
extended simultaneous observations of Markarian 421 over the optical,
UV, GeV, and TeV energy regime are crucial to establish the correlations
of non-thermal photon emission with the ultimate goal to understand
the gamma-ray production and emission process in blazars.

\acknowledgments

We thank Kevin Harris and Teresa Lappin for their considerable help in
obtaining the observations. The Whipple Collaboration acknowledges
supported from the US Department of Energy, NASA, the Smithsonian
Scholarly Studies Research Fund, and EOLAS, the scientific funding
agency of Ireland. S.B. acknowledges the financial support of the
National Science Foundation.

\pagebreak

\pagebreak

\onecolumn

{\small
\begin{center}
TABLE 1

{\sc Instrument Parameter}

\smallskip
\begin{tabular}{llccc} \hline
\hline
Season		&Mode		& E$_{Th}$	& Collection Area	&Trigger	\\
		&		& [GeV]		&[cm$^2$]		&		\\ \hline
1991$-$1992     &on/off		& 500		& 3.5$\times 10^8$ 	& m=2,n=50	\\
1992$-$1993     &on/off		& 400		& 2.2$\times 10^8$ 	& m=3,n=36	\\
1993$-$1994     &on/off, tracking& 350		& 3.5$\times 10^8$	& m=2,n=30-40	\\
1993$-$1994 (LC)&on/off, tracking& 250		& 3.5$\times 10^8$ 	& m=2,n=30-40	\\ \hline
\end{tabular}

\vglue 0.5in
TABLE 2

{\sc Observation Summary}

\smallskip
\begin{tabular}{lclccc} \hline
\hline
Observation Period		&$E_{Th}$	&Mode 		&Duration	&Nights	& Gamma-Ray Rate	\\ 
				&[GeV]		&		&[min]		&	& [min$^{-1}$]		\\ \hline
Mar 24, 1992 - Jun 2, 1992	& 500   	& on/off	&445	  	& 8    	& 0.33$\pm$0.05\\
Dec 23, 1992 - Feb 26,1993	& 400		& on/off	&750		& 10   	& 0.29$\pm$0.08\\
Dec 15, 1993 - May 9, 1994 	& 350		& on/off	&1145 		& 16   	& 0.31$\pm$0.06\\
Jan 10, 1994 - May 3, 1994	& 350   	& tracking	&939		& 9  	& 0.17$\pm$0.06\\ 
Jan 17, 1994 - June 12, 1994	& 250   	& on/off	&665		& 15  	& 0.88$\pm$0.14\\ 
Mar  5, 1994 - May 30, 1994	& 250   	& tracking	&516		& 5  	& 1.47$\pm$0.20\\ \hline
\end{tabular}

\vglue 0.5in
TABLE 3

{\sc Average Season Fluxes}

\smallskip
\begin{tabular}{lcc} \hline
\hline
Observing Season    &$E_{Th}$[GeV] & Flux $ [cm^{-2}s^{-1}]$\\ \hline
1991/1992           & 500          & 1.59$\pm$0.20 $\times 10^{-11}$ \\
1992/1993           & 500          & 1.47$\pm$0.30 $\times 10^{-11}$ \\
1993/1994 pre       & 350          & 1.18$\pm$0.20 $\times 10^{-11}$ \\
1993/1994 burst     & 250          & 21.90$\pm$3.80 $\times 10^{-11}$ \\
1993/1994 post      & 250          & 2.48$\pm$0.60 $\times 10^{-11}$ \\ \hline
\end{tabular}

\vglue 0.5in
TABLE 4

{\sc Probability for Constant Emission}

\smallskip
\smallskip
\begin{tabular}{lcrrc} \hline
\hline
Observing Season	& \multicolumn{3}{c}{Markarian 421}	& Crab Nebula \\ 
			& Prob.		& $\chi^2$	& ndf	& Prob. \\ \hline
1991/1992     		& 1.9 $\times$ 10$^{-2}$& 16.8&7&0.48        \\
1992/1993     		& 0.84                  & 5.0&9&0.66 \\
1993/1994 pre-burst     & 1.3 $\times$ 10$^{-9}$& 97.4&28&0.78 \\
1993/1994 post-burst    & 0.11           	& 15.6&10&no data \\ \hline
\end{tabular}
\end{center}
}

\pagebreak



\UGinsert{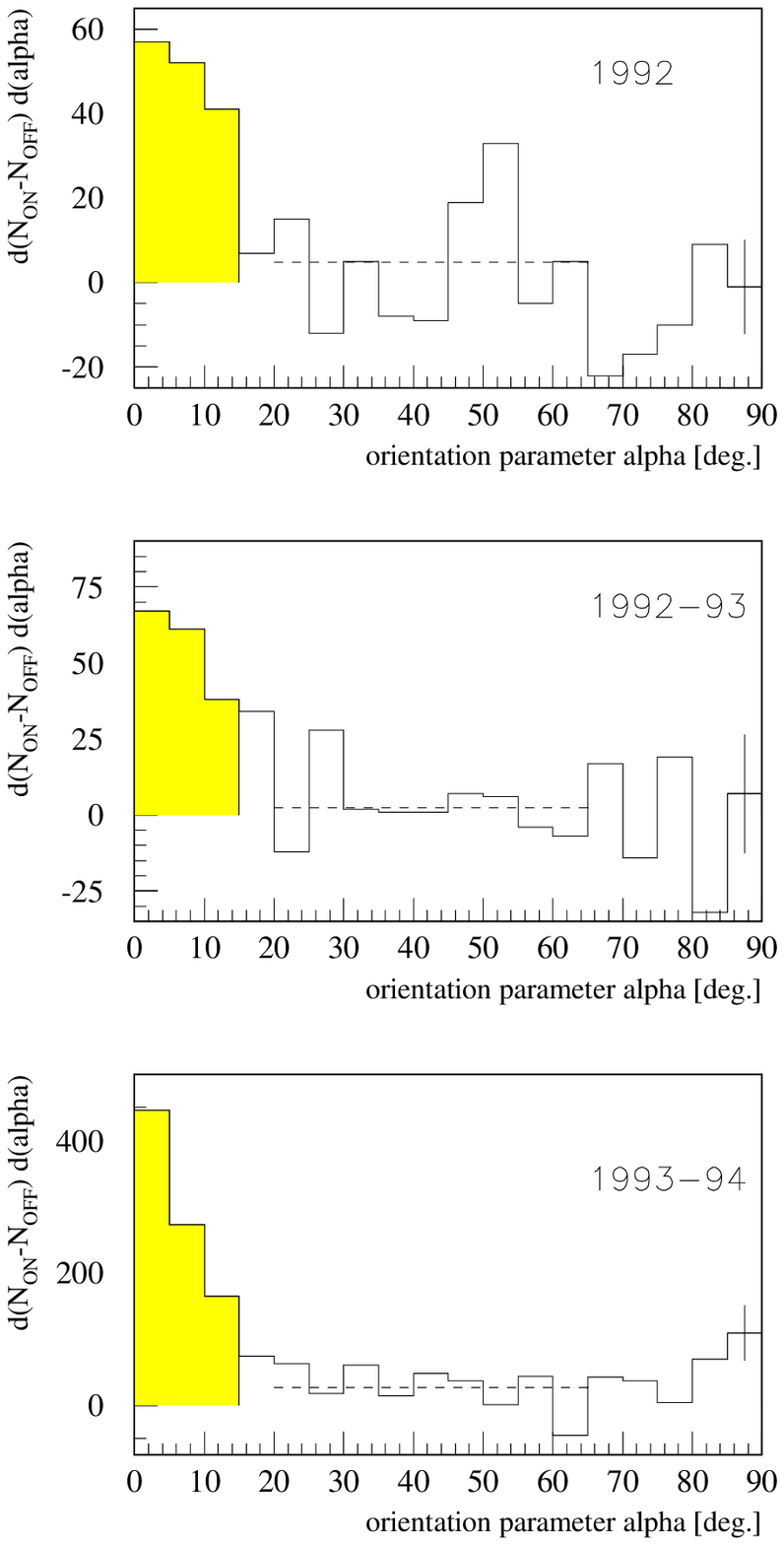}{6.5}{8.5}{.425}{.425}{0.7}

{\bf Figure 1.} On-source minus off-source orientation angle ($alpha$)
distributions for the three viewing periods 1992 (a), 1992/1993 (b), and
1993/1994 (c). The net excess in gamma-ray like events accepted by the
selection procedure is indicated
by the shaded area. The dashed line shows the average net gamma-ray rate
for images with large $alpha$ angles. This rate is expected to be zero.
A one sigma statistical error bar is always shown
for the bin 85$-$90\deg.

\pagebreak

\UGinsert{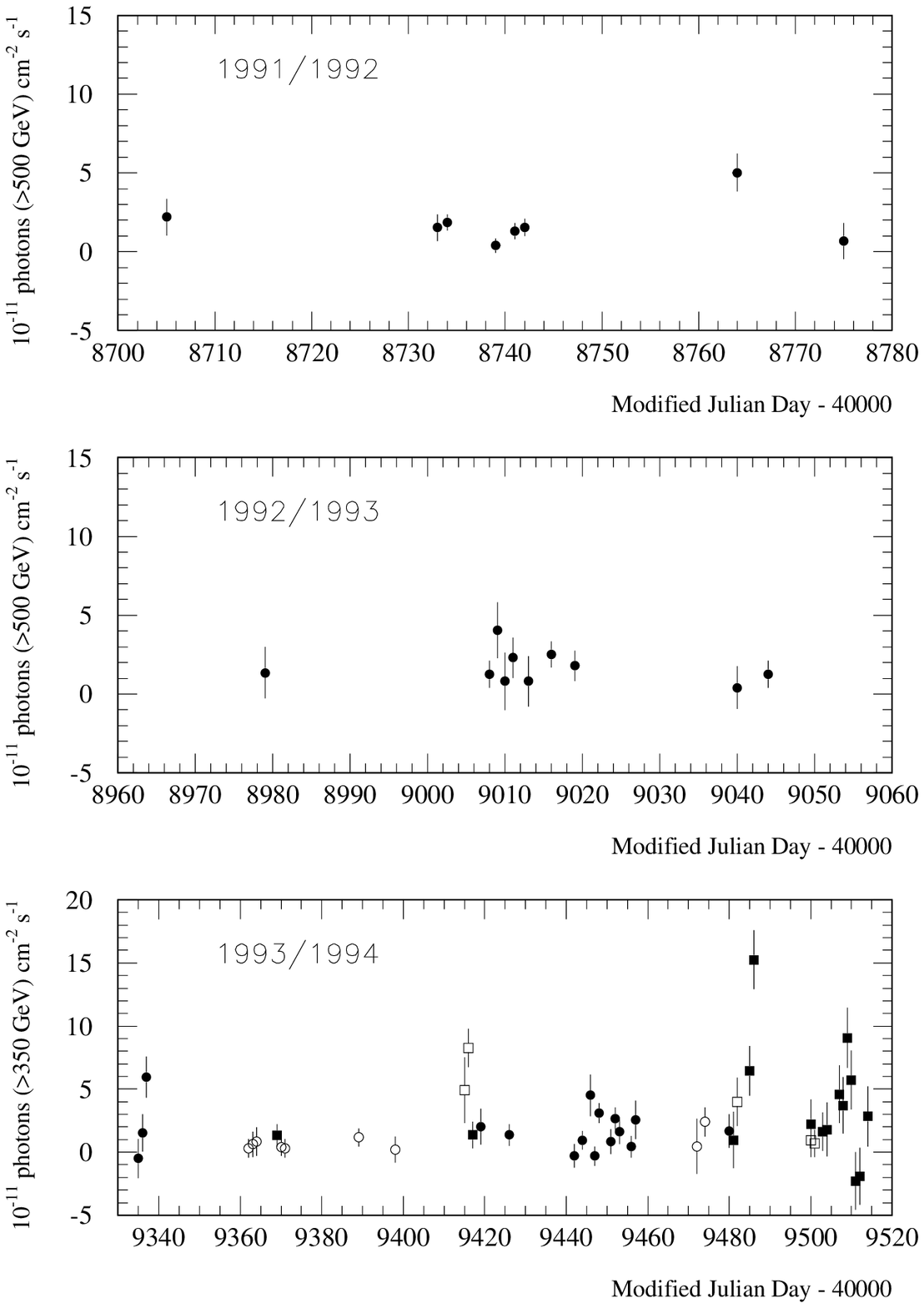}{6.5}{8.5}{.425}{.425}{0.7}

{\bf Figure 2.} The observed fluxes from Markarian 421 for
the three viewing seasons, (a) 1992, (b) 1993, and (c) 1994. Each data
point corresponds to one night of observation. Different symbols
indicate different instrument parameter and operation mode: solid
circle - on/off mode, no light cones; open circle - tracking mode, no
light cones; solid box - on/off mode, light cones; open box - tracking
mode, light cones.

\pagebreak

\UGinsert{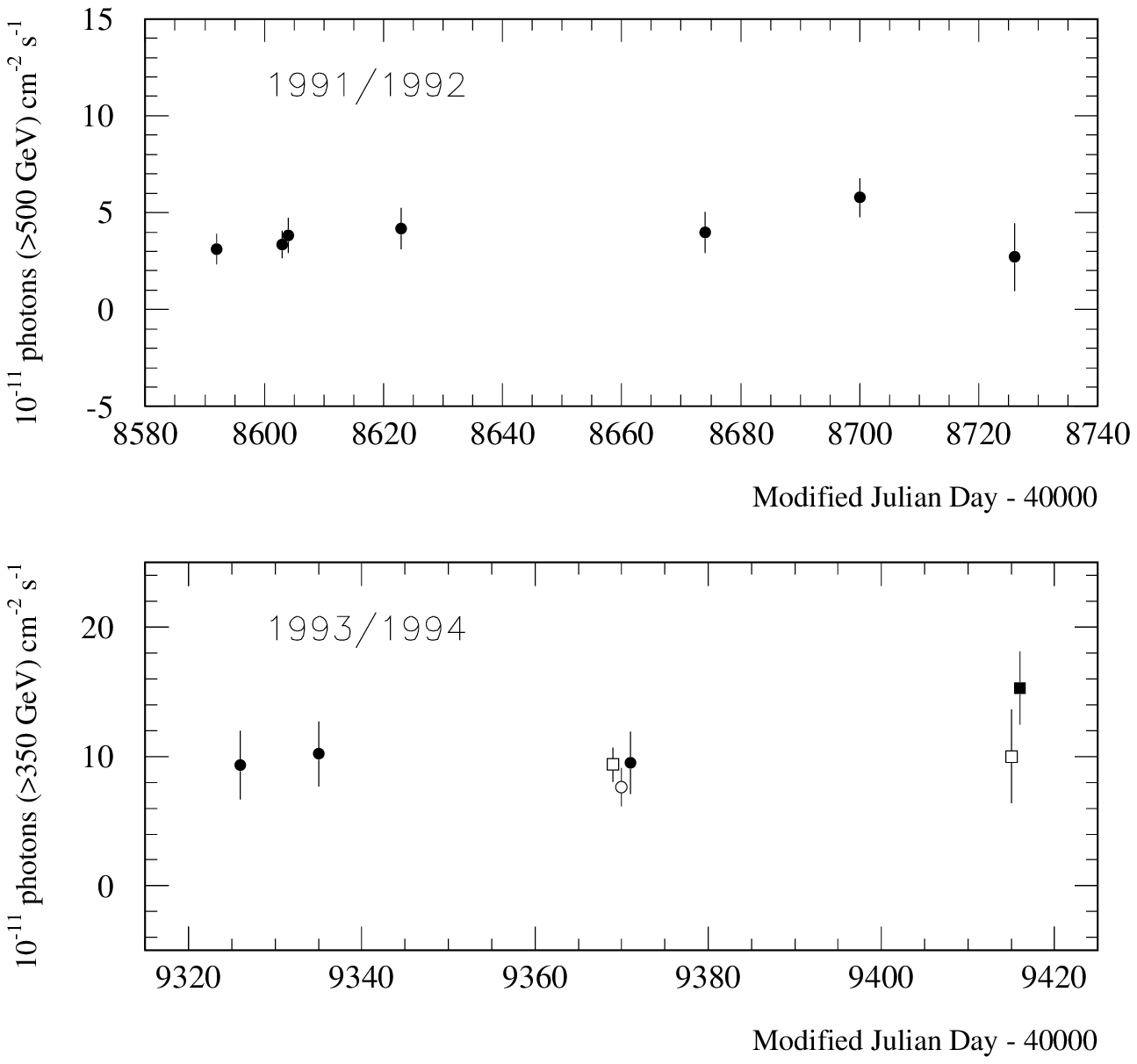}{6.5}{8.5}{.425}{.425}{0.7}

{\bf Figure 3.} The observed fluxes from the Crab Nebula for
the viewing seasons 1992 (a), and 1994 (b). Symbols as in Figure 2.

\pagebreak

\UGinsert{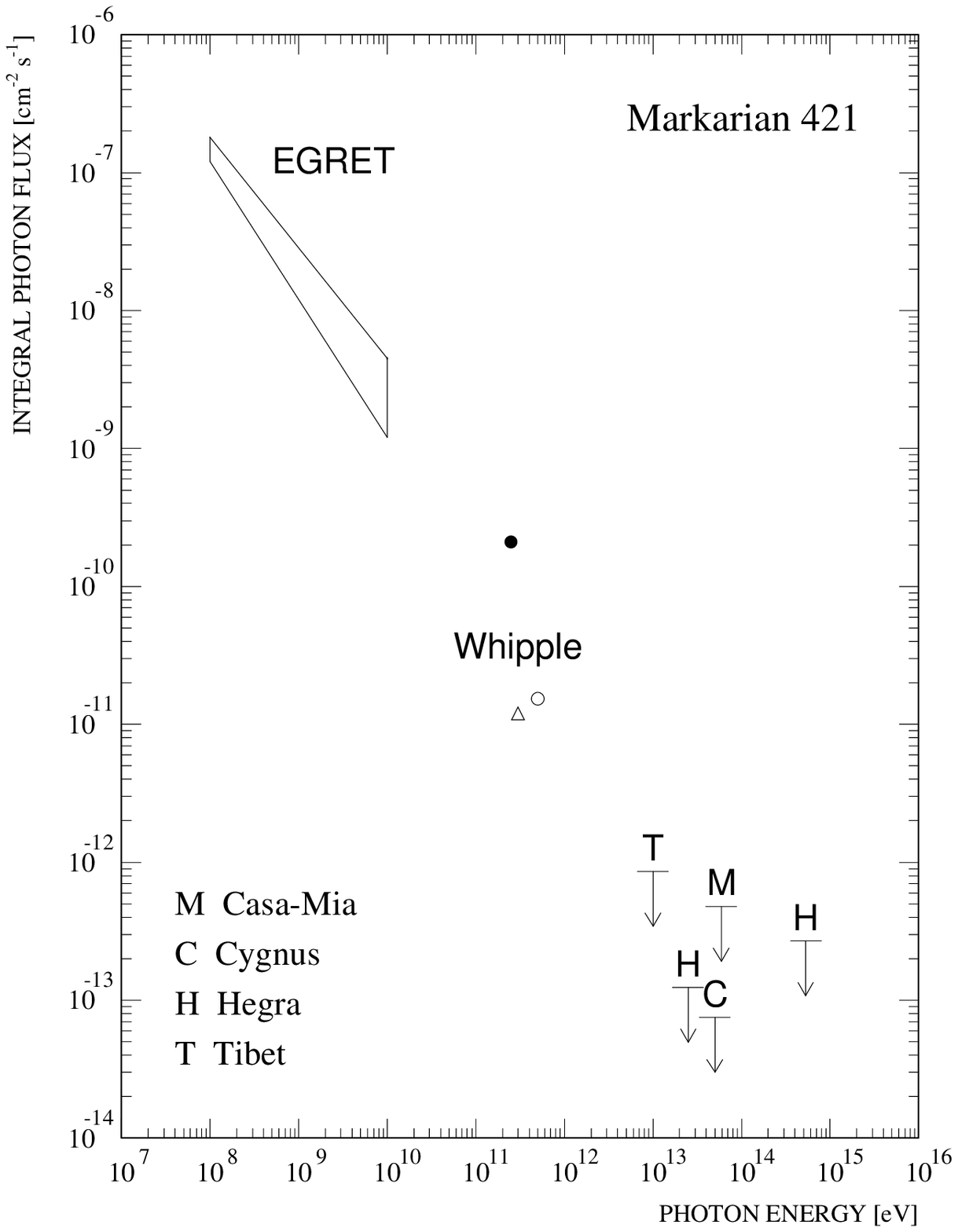}{6.5}{8.5}{.425}{.425}{0.7}

{\bf Figure 4.} Integral photon spectrum for Markarian 421. The error box
for EGRET was converted from the differential fit to the observed data
(Lin et al. 1993). Upper limits are from air shower experiments: HEGRA
(Karle 1994; K\"uhn 1994); Cygnus (Alexandreas et al. 1993); Tibet
(Amenomori et al. 1994); CASA-MIA (Catanese et al. 1994).  The Whipple
data points are the average integrated flux for 1992 and 1992/1993
(open circle), the pre-burst average flux in 1994 (open triangle), and
the flux observed during the burst in May 1994 (solid circle).

\end{document}